\documentclass[5p,twocolumn]{elsarticle}
\usepackage{graphicx,latexsym}
\usepackage{dcolumn}
\usepackage{amssymb,amsmath,bm}
\usepackage{subfigure}
\usepackage{braket}
\usepackage{siunitx}
\usepackage{array}
\usepackage{float}
\usepackage[nopar]{lipsum}
\usepackage{caption}
\usepackage{multirow}
\usepackage{bigstrut}
\usepackage[super]{nth}

\newcommand{\angstrom}{\text{\normalfont\AA}}
\usepackage{hyperref}
\hypersetup{
    pdfnewwindow=true,       
    colorlinks=true,         
    linkcolor=blue,          
    citecolor=blue,          
    filecolor=magenta,       
    urlcolor=black           
}

\usepackage[normalem]{ulem}

\def\sec#1{Sec.\ \ref{#1}}

\def\fig#1{Fig.\ \ref{#1}}
\def\tab#1{Tab.\ \ref{#1}}

\journal{}

\begin{document}

\begin{frontmatter}



\title{Interaction effects in a two-dimensional AlSi$_6$P nanosheet: A first-principles study on the electronic, mechanical, thermal, and optical properties}

\author[a1,a2]{Nzar Rauf Abdullah}
\ead{nzar.r.abdullah@gmail.com}
\address[a1]{Division of Computational Nanoscience, Physics Department, College of Science, University of Sulaimani, Sulaimani 46001, Kurdistan Region, Iraq}
\address[a2]{Computer Engineering Department, College of Engineering, Komar University of Science and Technology, Sulaimani 46001, Kurdistan Region, Iraq}

\author[a1]{Hunar Omar Rashid}

\author[a3]{Andrei Manolescu}
\address[a3]{Reykjavik University, School of Science and Engineering, Menntavegur 1, IS-101 Reykjavik, Iceland}

\author[a4]{Vidar Gudmundsson}
\address[a4]{Science Institute, University of Iceland, Dunhaga 3, IS-107 Reykjavik, Iceland}


\begin{abstract}

Physical quilities such as electronic, mechanical, thermal, and optical properties of Al and P codoped silicene forming AlSi$_6$P nanosheets are investigated by first-principle calculations within density functional theory. A particular attention of this study is paid to the interaction effect between the Al and P atoms in the buckled silicene structure. We infer that the interaction type is attractive and it leads to an asymmetry in the density of states opening up a band gap. In contrast to BN-codoped silicene, the buckling in the Al-P codoped silicene is not much influenced by the dopants as the bond lengths of Si-P and Al-P are very similar to the Si-Si bond lengths. 
On the other hand, the longer bond length of Si-Al decreases the stiffness and thus induces fractures at smaller values of applied strain in AlSi$_6$P. The elastic and nonelastic regions of the stress-strain curve of AlSi$_6$P depend on the placement of the Al and P atoms.
The finite band gap caused by the Al-P dopant leads to enhancement of the Seebeck coefficient and the figure of merit, and induces a redshift of peaks in the dielectric response, and the optical conductivity. Finally, properties of the real and imaginary parts of the dielectric function, the excitation spectra, the refractive index, and the optical response of AlSi$_6$P are reported.   
\end{abstract}


\end{frontmatter}

\section{Introduction}

There has been a rapid development of two-dimensional (2D) material structures such as silicene \cite{LeLay2015}, graphene \cite{Ren2014}, molybdenum disulfide (MoS$_2$) \cite{Kibsgaard2012} and hexagonal boron nitride \cite{Zhang_2017} in several fields owing to their promising electrical, thermal, mechanical, and optical properties with potential applications in opto- and nanoelectronics \cite{Tao2015}. The stability of these 2D structures plays an important role in the investigation of their physical and chemical properties \cite{Kharadi_2020}.
It has been seen that the buckled structure of 2D-silicene is indeed thermodynamically stable \cite{PhysRevLett.102.236804}.
The breaking of the mirror symmetry in buckled silicene eliminates instabilities associated with the high-symmetry structures.
The stability of 2D-silicene structures leads to very attractive physical properties for a wide range of applications such as in nano-electronics in silicon-based technology \cite{OUGHADDOU201546, RASHID2019102625}.

The semi-metallic nature of 2D silicene is not 
so suitable for nano-electronic devices. For practical applications a material with a finite band gap is usually preferred. 
A finite band gap in silicene can be created by structural defects \cite{Li2015}, or by applying uniaxial strain \cite{Houmad2019}, or an electric field \cite{PhysRevB.85.075423}. In contrast to graphene, 
the advantage of having a buckled structure in silicene is that the band gap can be tuned by
applying a transverse external static electric field \cite{doi:10.1021/nl203065e}, and
the size of band gap in silicene increases linearly with the electric field strength.

As an alternative way, researchers recently focused on 2D-silicene materials with impurities of B-N dopants \cite{ABDULLAH2020114556}, or by using Al-P doping \cite{C4RA07976K}, their physical properties  \cite{LI2021150041, BAFEKRY2020113850, ABDULLAH2021114644} indicating the potential of silicene in different fields. 
Similar to B and N doping in graphene \cite{ABDULLAH2020126807}, good dopants for 
silicene are Aluminium (Al) and Phosphorous (P). The Al-P codoped silicene forming AlSi$_x$P structures has two advantages. First: the lattice deformation is
small due to the P and Al dopant atoms, because the atomic radii of Al
and P are close to that of a Si atom \cite{PhysRevB.87.085444}. Second: the system remains iso-electronic which is due to the presence of buckling in silicene and thus one may expect higher reactivity towards Al and P atoms. This strong reactivity is confirmed by the first-principles computations with the generalized gradient approximation (GGA) in the framework of PBE potential. It has been shown that Al and P can interact strongly with the Si atoms in silicene \cite{HERNANDEZCOCOLETZI2018242} which gives rise a suitable binding energy to form 
the AlSi$_x$P structures. The study of electron energy loss function of AlSi$_x$P structure have shown that 
two new small, but important electron energy loss spectra (EELS) peaks emerge for P doping in silicene in the case of perpendicular polarized electric field, while no new peak is found irrespective of the dopant type for the parallel polarization. The reason for these new EELS peaks has been referred to the inherent buckling effect 
of stable silicene \cite{C4RA07976K}.

 \lipsum[0]
\begin{figure*}[htb]
	\centering
	\includegraphics[width=1.0\textwidth]{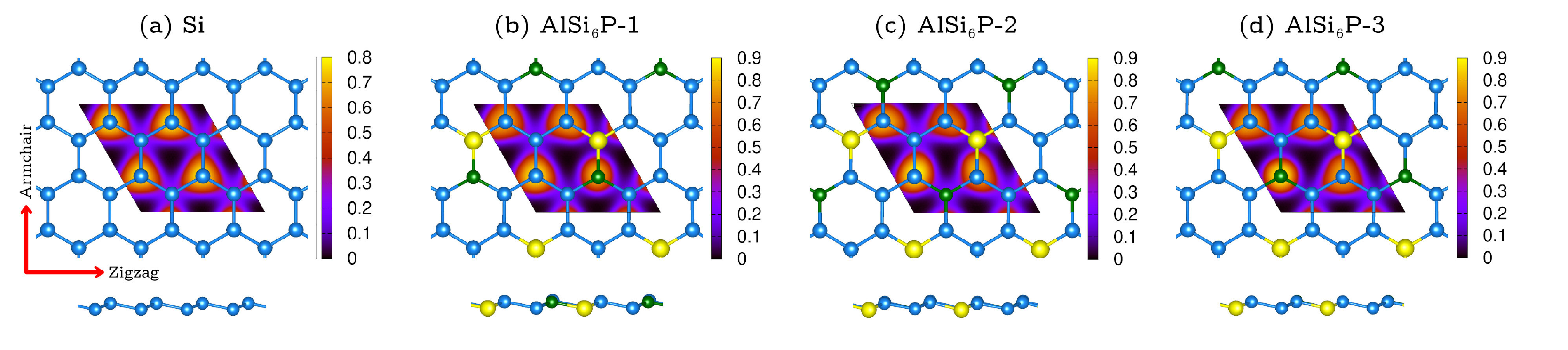}
	\caption{Pristine silicene (Si) nanosheet (a), AlSi$_6$P-1 (b), AlSi$_6$P-2 (c) and AlSi$_6$P-3 (d) nanosheets were the Si atoms are blue,
		Al atoms are yellow, and P atoms are green. The contour plots indicate the electron localization function (ELF). The Al atoms are doped at ortho-position in all AlSi$_6$P structures while the P atom is doped at meta-, and para- and meta-position in AlSi$_6$P-1, AlSi$_6$P-2, and AlSi$_6$P-3, respectively.  The bottom panel is the side view of the corresponding structures.}
	\label{fig01}
\end{figure*}

In this work, motivated by the route undertaken by
recent studies of silicene, we investigate
the electronic, mechanical, thermal and optical properties of silicene doped with P and Al atoms.
Our interest is pointed towards the interaction effect between the dopant atoms on the physical properties of 
Al-P codoped silicene. We have found that different atomic configurations of Al and P atoms in silicene 
monolayer influence the interactions between the dopant atoms leading to significant changes in the physical properties of the different AlSi$_x$P structures.

In \sec{Sec:Model} the silicene structure is briefly overviewed. In \sec{Sec:Results} the main achieved results are analyzed. In \sec{Sec:Conclusion} the conclusion of results is presented.

\section{Computational Tools}\label{Sec:Model}

In this work, the Al-P codoped silicene monolayers are studied with the method of density functional theory (DFT) by solving the standard Kohn-Sham equations as 
implemented in the Quantum Espresso (QE) package \cite{Giannozzi_2009, giannozzi2017advanced}.
Crystalline and molecular structure visualization software (XCrySDen) and VESTA
are used to visualize the structures studied in this work \cite{KOKALJ1999176, momma2011vesta}.
After modeling pristine silicene and the AlSi$_6$P structures, 
the generalized gradient approximation (GGA) with the Perdew-Burke-Ernzerhof (PBE) functionals \cite{PhysRevLett.77.3865} is employed to study the physical properties of AlSi$_6$P.

A Monkhorst-Pack grid point mesh of size $14\times14\times1$ is used for the Brillouin-Zone (BZ) sampling integration \cite{PhysRevB.13.5188,ABDULLAH2021106073}. After checking convergence with respect to an energy cutoff, we find that a value of $1088.45$ eV leads to well converged results. 
Using the grid-point mesh and the energy cutoff values, the structures are optimized
by minimizing the forces on individual atoms below  $10^{-5}$ eV/$\angstrom$.
The distance between periodically repeated images of the different systems along the $z$-direction is set to at least $20$~$\angstrom$ \cite{ABDULLAH2021106981}.

The grid point mesh utilized to calculate the density of state (DOS) is $100\times100\times1$. 
Once the DOS is obtained one can use a Boltzmann transport theory (BoltzTraP) to investigate the thermoelectric behavior of the systems \cite{madsen2006boltztrap-2}. The BoltzTraP code employs a mesh of band energies and is interfaced to the QE package \cite{ABDULLAH2020126578}. 
The optical properties of the systems are evaluated by the QE code, and 
an optical broadening of $0.1$~eV is considered for the calculation of
the dielectric properties.

\section{Results}\label{Sec:Results}

In this section, we present the results for pristine silicene and Al-P codoped silicene. 
A 2 × 2 supercell of buckled silicene structure is modeled in the calculation.
In addition to the pristine silicene structure (a), we consider different Al-P codoping schemes of silicene to make three configurations of AlSi$_6$P (b-d) as are shown in \fig{fig01}: Al is fixed at the ortho-position (yellow) in all structures, and the P atom (green) is assumed to be at meta-, para-, meta-position with respect to the Al atom leading to structures to be identified as AlSi$_6$P-1 (b), AlSi$_6$P-2 (c), AlSi$_6$P-3 (d), respectively.
The buckling of pristine silicene shown in the side view (bottom panel) is caused by the positions of the Si atoms on the A and B sites that are displaced out of the plane leading to
a staggered sublattice potential and a layer separation between the two sublattices of the A and B sites.
One notices that the distance between the Al and P atoms in these structures are different leading to different interaction strength between them, which should affect the physical properties of the AlSi$_6$P structures as will be been seen later.

After the structure relaxation, the lattice constant, $a$, the buckling length, $\delta$, and the bond lengths for the pure silicene and AlSi$_6$P structures are calculated and presented in \tab{table_one}. 
\begin{table}[h]
	\centering
	\begin{center}
		\caption{\label{table_one} Lattice constant, a, buckling length, $\delta$, Si-Si, Al-P, Si-Al, and Si-P bonds for all pure silicene and Al-P codoped structures. The unit of all parameters is $\angstrom$.}
		\begin{tabular}{|l|l|l|l|l|l|l|}\hline
			Structure	  & a     & $\delta$& Si-Si & Al-P  & Si-Al & Si-P   \\ \hline
			Si	          & 3.86  &  0.45   & 2.27  & -     & -     & -       \\
			AlSi$_6$P-1	  & 3.88  &  0.449  & 2.28  & 2.312 & 2.353 & 2.239   \\
			AlSi$_6$P-2	  & 3.90  &  0.485  & 2.288 & -     & 2.341 & 2.271   \\
			AlSi$_6$P-3	  & 3.903 &  0.491  & 2.295 & -     & 2.329 & 2.309   \\ \hline
		\end{tabular}
	\end{center}
\end{table}
The values of $a$, $\delta$, and the Si-Si bond length of pure silicene are in a good agreement with previous studies for slightly buckled silicene \cite{PhysRevLett.102.236804}. The value of the bond angle is $\approx 116^{\circ}$ indicating the buckling structure of pure silicene while the bond angle of planar silicene is $\approx 120^{\circ}$ \cite{C4RA04174G}. In addition the values of $a$, $\delta$, and Si-Si, Al-P, Si-Al, and Si-P are found to be in agreement with results in the literature \cite{C4RA07976K}.

\subsection{Interaction energy}

The interaction energy between the the Al and P dopant atoms in
silicene is determined from the total energy of the system using a DFT calculation. 
The interaction energy between the Al and P dopant atoms in a structure is obtained via \cite{ABDULLAH2020100740}
\begin{equation}
	\Delta E = E_2 - E_0 + 2 \times E_1,
\end{equation}
where $E_0$, $E_1$ and $E_2$ are the total energies of the systems with zero, one
and two substitutional dopant atoms, respectively. We observe that the interaction is attractive as the interaction energy has a negative value for all the three considered configurations of AlSi$_6$P.
The interaction energies for AlSi$_6$P-1, AlSi$_6$P-2, AlSi$_6$P-3 are found to be $-3.17$, $-4.38$, and $-5.14$~eV, respectively, revealing a stronger attractive interaction between the two substitutional Al and P atoms in the AlSi$_6$P-1 structure. This is expected as the distance between the Al and P atoms is shorter in AlSi$_6$P-1 compared to the other two AlSi$_6$P structures.
The interaction between the Al and P atoms influences the geometry of structures. For instance, the value of $a$ and $\delta$ of AlSi$_6$P-1 are almost the same as in pure silicene, while they are increased with decreasing attractive interaction between the Al and P atoms in AlSi$_6$P-2, and AlSi$_6$P-3 (see \tab{table_one}).

\subsection{Structural stability}

We study the formation energy of all the structures. Formation energy is the energy required for generating the atomic configuration of the structure, which indicates the energetic stability of doped silicene structures. The formation energy can be defined as
\begin{equation}
E_f = E_T - N_{\rm Si} \, \mu_{\rm Si} - \sum_i N_i \, \mu_i, 
\end{equation} 
where $E_{\rm T}$ is the total energy of the doped silicene system, $N_{\rm Si}$ and $\mu_{\rm Si}$ refer to the number and chemical potential of the Si atoms, respectively, and $N_{\rm i}$ and $\mu_{\rm i}$ are the number and the chemical potential of the doped atoms (Al and P atoms), respectively.
We have found that the formation energy of AlSi$_6$P-1, AlSi$_6$P-2, AlSi$_6$P-3 are 
$-35.9405$, $-35.4192$, $-35.3825$~eV, respectively.
We expect that, the smaller the formation energy, the more energeticly stable the structure should be. Consequently, the AlSi$_6$P-1 is more energeticly stable than the  AlSi$_6$P-2 and AlSi$_6$P-3 structures. It should be pointed out that the formation energy
is calculated from total energy calculations and does not
take into account dynamic terms.
In addition, the binding energy calculations of Al-P codoped silicene indicate that 
the substitutional doping is energetically more favorable for P atoms and it is less favorable for Al atoms. This can be connected to the bond lengths of Si-P which is sligltly shorter than the Si-Si bonds, but the Si-Al bond length is larger than the 
Si-Si bond (see \tab{table_one})\cite{HERNANDEZCOCOLETZI2018242}.

\subsection{Band structures}

The Si atoms have four valence electrons, and the Al and P atoms have 
three and five valence electrons, respectively. The concentration of Al atom increases the hole concentration resulting in a p-type doping of silicene, while the P atom increases the electron concentration leading to an n-type doping of silicene. We can therefore see from the electron localization function in \fig{fig01} that the electron density around the Al atom is low while a high electron density around the P atoms is found. 
In the vicinity of the Fermi energy, the Al atoms move the Fermi level inside the valence band of silicene, while the P atoms shift the Fermi level to the conduction band \cite{HERNANDEZCOCOLETZI2018242}. But the Fermi energy of Al-P codoped silicene stays between the valence and the conduction bands, which is caused by the combined effects of the Al and P atoms leading to a semiconducting material.  
\begin{figure}[htb]
	\centering
	\includegraphics[width=0.4\textwidth]{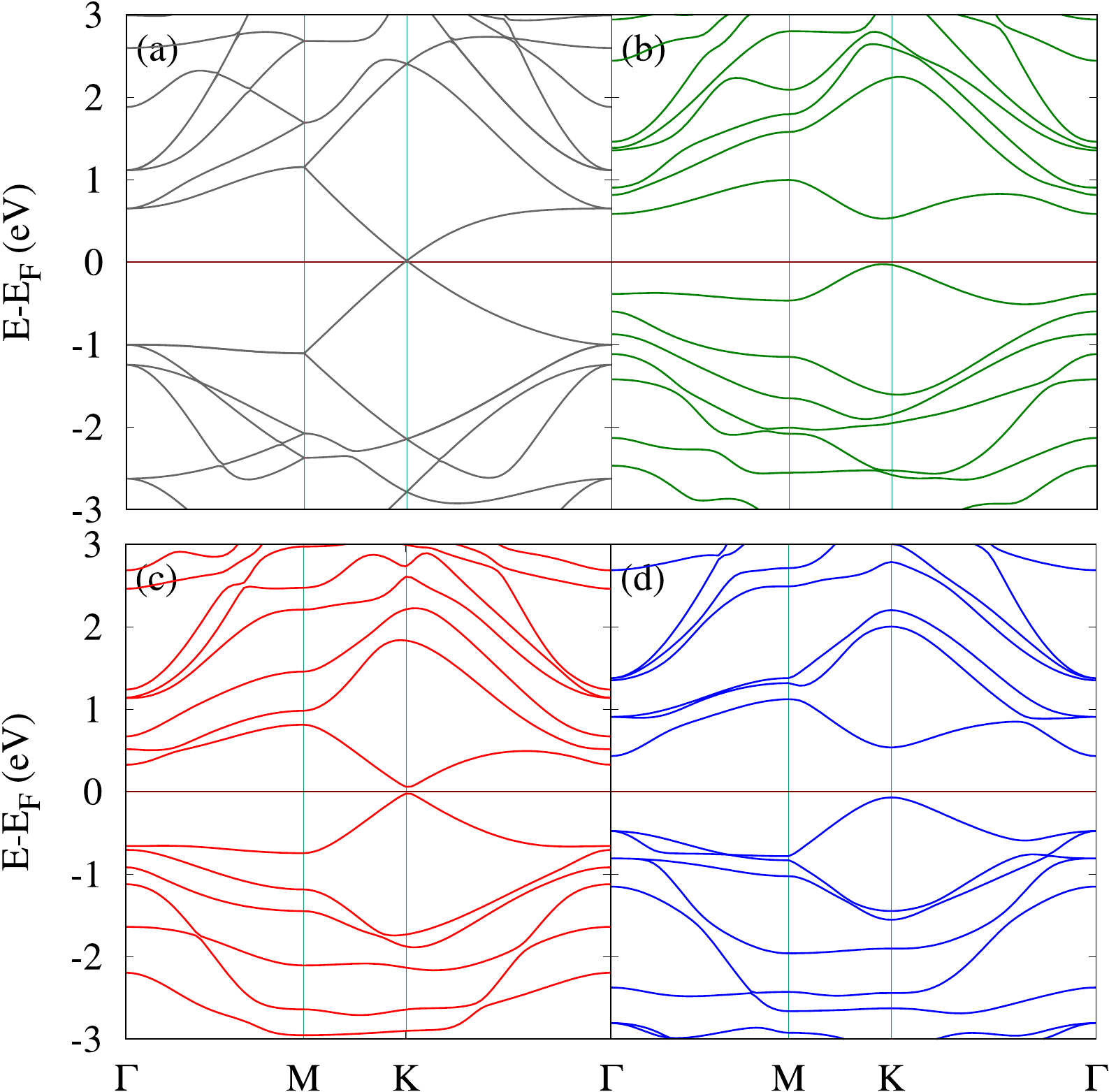}
	\caption{The electronic band structure of pure Silicene (a), AlSi$_6$P-1 (b), AlSi$_6$P-2 (c), and AlSi$_6$P-3 (d).  Fermi energy is set at 0 eV. }
	\label{fig02}
\end{figure}
The electronic band structure is plotted in \fig{fig02} for pure silicene (a), AlSi$_6$P-1 (b), AlSi$_6$P-2 (c), and AlSi$_6$P-3 (d) along the high symmetry lines
in the Brillouin zone. Similar to graphene, the valence band maxima ($\pi$ band) and the conduction band minima ($\pi^*$ band) of pure silicene cross at the K point, which gives rise to a semimetal with a zero electronic band gap and linearly crossing bands at the Fermi level \cite{ABDULLAH2021114644}.
The linear relation is due to the zero value of the onsite energy (energy difference) between the Si atoms at the A and B sites of the hexagon of pure silicene arsing from the inversion symmetry of the structure \cite{aliofkhazraei2016graphene}.

In the three configurations of AlSi$_6$P structures, the Al and P atoms are either doped at the A and B sites or A(B) site. In AlSi$_6$P-1 and AlSi$_6$P-3, the Al and P atoms are situated at A and B sites, respectively, while in the AlSi$_6$P-2 both Al and P atoms are put at A or B sites. Consequently, 
the onsite energy of the AlSi$_6$P-1 and AlSi$_6$P-3 are higher than those of AlSi$_6$P-2 arsing from a higher breaking of the inversion symmetry of the AlSi$_6$P-1 and AlSi$_6$P-3 structures. 
The band gap of AlSi$_6$P-1, $0.552$ eV, and AlSi$_6$P-3, $0.606$ eV, are thus higher than that of AlSi$_6$P-2, $0.083$ eV.
In addition, the linear dispersion of the energies around the Fermi energy has vanished in both  
the AlSi$_6$P-1 and AlSi$_6$P-3, which is again due to the breaking of the inversion symmetry.

\subsection{Density of States}

\begin{figure}[htb]
	\centering
	\includegraphics[width=0.45\textwidth]{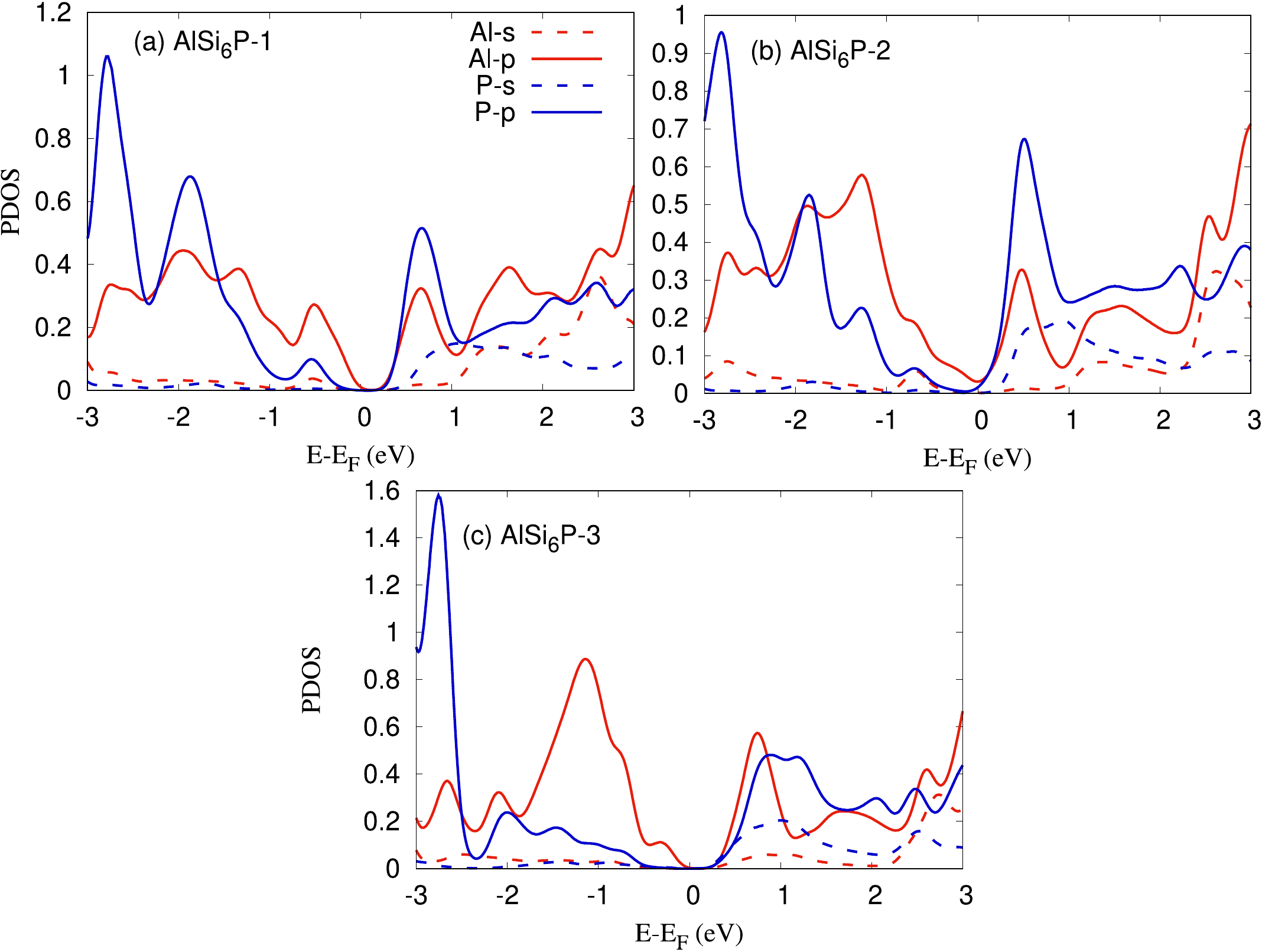}
	\caption{Partial density of state (PDOS) of AlSi$_6$P-1 (a), AlSi$_6$P-2 (b), and AlSi$_6$P-3 (c). Fermi energy is set at 0 eV.}
	\label{fig03}
\end{figure}

We present the partial density of states (PDOS) analysis for AlSi$_6$P-1 (a), AlSi$_6$P-2 (b), and AlSi$_6$P-3 (c) in \fig{fig03}, where the contributions of $s$ (dashed line) and $p$ (solid line) orbitals for Al (red) and P (blue) atoms in the PDOS are shown. 
The major $p$ orbital of both the Al and P atoms are found near the Fermi level, while the major contribution of the $s$ orbitals of both atoms are seen at the energy $>0.5$~eV. In addition, the participation of the Al atoms in the PDOS of valence band is higher than those of the P atom around the Fermi energy, which confirms that the Al atoms lead to p-type semiconduction. In contrast, the PDOS of the P atom at the conduction band near the Fermi level is higher than those of the Al atom. It is interesting to see that the attractive interaction between the Al and P atoms influences and tunes the contribution of the Al and P atoms in the PDOS in both the valence and the conduction band regions. 
In the presence of a strong attractive interaction between the Al and P atoms of AlSi$_6$P-1, a balance between the $p$-orbitals of both atoms near the Fermi energy is observed. With decreasing attractive interaction in AlSi$_6$P-2 and AlSi$_6$P-3, the contribution of the Al atom in the PDOS of valence band near the Fermi energy is enhanced, while the contribution of the P atom in the conduction band is almost unchanged. The contribution of the Al and P atoms breaks the symmetry of the density of states around the Fermi energy influencing the physical properties of the systems.

\subsection{Mechanical properties}

\begin{figure}[htb]
	\centering
	\includegraphics[width=0.5\textwidth]{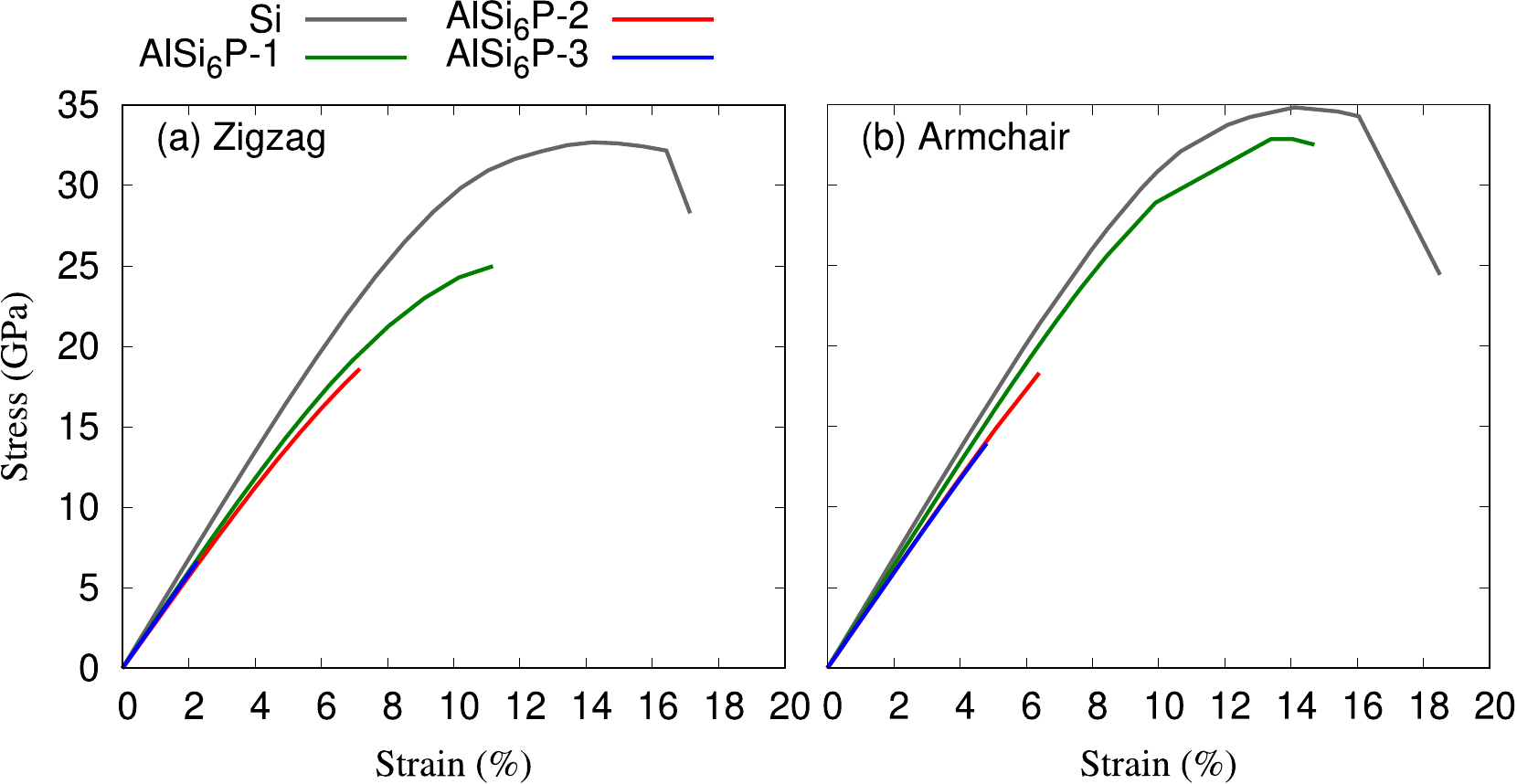}
	\caption{Uniaxial stress-strain curves of pure silicene, and AlSi$_6$P monolayers along 
		     the zigzag (a) and armchair (b) directions.}
	\label{fig04}
\end{figure}

We examine the mechanical properties of pure silicene and AlSi$_6$P structures by conducting a uniaxial tensile simulation \cite{MORTAZAVI2017228}. Commonly, we calculate the mechanical responses along the armchair and zigzag direction as it is shown in \fig{fig04}. Likely to 2D graphene, the pure silicene is found to exhibit isotropic elasticity which means that the elastic modulus along the zigzag and armchair directions are the same. One can guess the mechanical behavior of a system from the bond lengths before calculating the stress-strain curve. The calculated bond lengths shown in \tab{table_one} indicate that the Si-Si bond length of pure silicene is smaller than that of Si-Si, and Si-Al, and almost equal to the Si-P bond lengths of AlSi$_6$P structures implying that pure silicene is mechanically harder than AlSi$_6$P structures. This observation is also reflected in the lattice constant which increases from 3.86~$\angstrom$ for pure silicene to 3.903~$\angstrom$ for AlSi$_6$P-3.       
It should be noticed that not only the bonds become weaker in AlSi$_6$P, but also the volume of it increases, resulting in substantially weaker mechanical properties.

The stress-strian curve of pure silicene and AlSi$_6$P structures are shown in \fig{fig04}. 
In the stress-strain curve calculations of pure silicene, we realize that the buckling length of the sheets are gradually decreased by increasing the strain level. An increase in the in-plane stiffness and elastic behavior are found from the linear relation between the stress and strain at the low values of strain ($\le 8 \%$) for both the zigzag and the armchair directions indicating isotropic elastic property. The ultimate strength of pure silicene is found to be $32.67$ GPa at strain $14.2\%$ in the zigzag, and $34.83$ GPa at strain $14.1 \%$ in the armchair directions. The ultimate strength is the maximum stress that the material can withstand while being stretched or pulled before breaking.    
As we have mentioned before, silicene is a relatively weak structure
compared to other 2D materials such as graphene due to
its buckled geometry. We therefore expect that the ultimate strength of silicene is lower than that of graphene \cite{ABDULLAH2020126350}. 
The elastic modulus is a quantity that measures a substance's resistance to being deformed elastically. The elastic modulus of pure silicene is found to be 333.33 for both the zigzag and the armchair directions.

In the AlSi$_6$P structures, the interaction between the Al and P atoms and the Si-Si, Al-P, Si-Al, Si-P bonds play an essential role for the stress-strain curve. As can be seen from \fig{fig04} the 
ultimate strength and the elastic modulus of AlSi$_6$P decrease with decreasing attractive interaction between the Al and P atoms from the AlSi$_6$P-1 to the AlSi$_6$P-3 structure. 
The elastic behavior of AlSi$_6$P is found from
the linear relation between the stress and strain at the low values of strain $4.8\%$, $4.03\%$, and $2.27\%$ in the zigzag direction, and $5.8\%$, $5.19\%$, and $4.81\%$ in the armchair direction for 
AlSi$_6$P-1, AlSi$_6$P-2, and AlSi$_6$P-3, respectively.
Furthermore, the ultimate strengths of AlSi$_6$P-1, AlSi$_6$P-2, and AlSi$_6$P-3 are found to be $24.98$, $18.6$, and $6.63$~GPa in the zigzag direction, respectively, and $32.86$, $18.33$, and $13.96$~GPa in the armchair direction, respectively. 

The first observation is that the AlSi$_6$P structures do no have isotropic elastic and ultimate strength properties. As we have mentioned, the strength of the attractive interaction between the Al and P atoms changes the Si-Si, Si-Al, and Si-P bond lengths resulting in different mechanical properties and non-isotropic elastic behaviors for the AlSi$_6$P structures. One can conclude that the mechanical properties of AlSi$_6$P are directly proportional to the strength of the attractive interaction between the Al and P atoms.

\subsection{Thermal properties}

In the next steps we study the thermoelectric properties of the AlSi$_6$P structures in the low temperature range, $20\text{-}165$~K, where the electron and the lattice temperatures are decoupled and the energy exchange between the charge carriers and the acoustic phonons is very weak \cite{PhysRevB.87.035415, ABDULLAH2021110095}. So, we need only to consider the electronic thermal properties in the selected range of temperature. To calculate the electronic part of thermal behavior one can utilize BoltzTraP.

\begin{figure}[htb]
	\centering
	\includegraphics[width=0.5\textwidth]{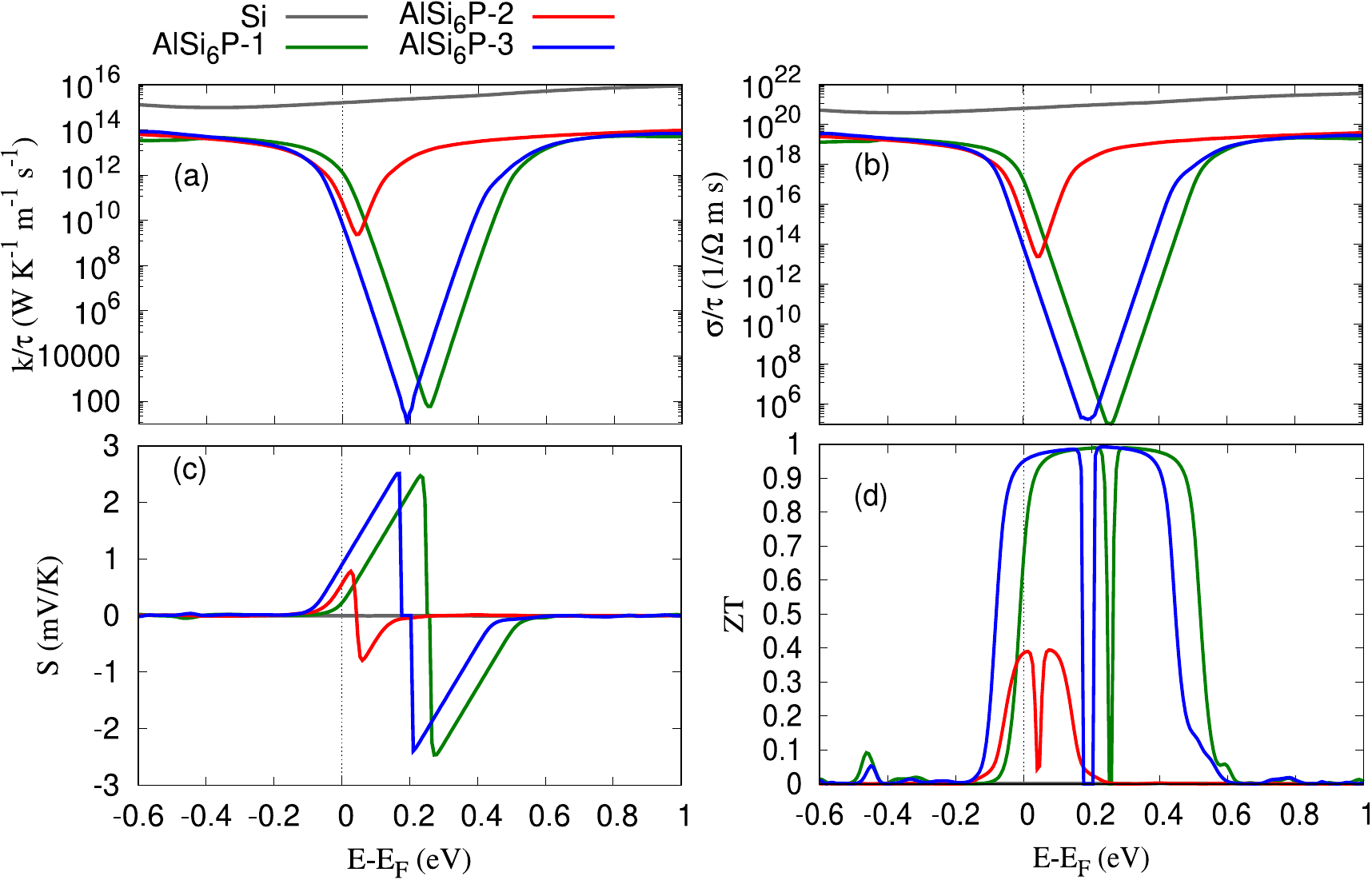}
	\caption{Electronic thermal conductivity, $k$ (a), electrical conductivity, $\sigma$ (b), Seebeck coefficient, $S$ (c), and figure of merit, $ZT$ (d) versus energy are plotted for pure silicene, Si, (gray), AlSi$_6$P-1 (green), AlSi$_6$P-1 (red), and AlSi$_6$P-1 (blue). The Fermi energy is set to zero.}
	\label{fig05}
\end{figure}

In thermoelectric devices, the parameters that can show thermoelectric ability of the device is the Figure of merit, $ZT$, or the power factor.
The $ZT$ is very well defined and it is $ZT = \sigma S^2 T/k$ where $\sigma$ is the electrical conductivity, $S$ refers to the Seebeck coefficient, $T$ is the temperature, and $k$ stands for the electronic part of the thermal conductivity \cite{ABDULLAH2021413273}. In order to obtain a good thermoelectric device or high $ZT$ at a specific temperature, one can assume a high $S$ and $\sigma$, and low $k$. 
To show the thermoelectric properties of the silicene and the AlSi$_6$P structures, we present \fig{fig05}, which shows $k$ (a), $\sigma$ (b), $S$ (c), and $ZT$ (d) versus energy. 
As a first glance, we can see that the pure silicene exhibits a high electronic thermal conductivity, and low Seebeck coefficient leading to a very low figure of merit (gray line). 

The thermoelectric properties of AlSi$_6$P show a sharp dip in the thermal conductivity indicating a low value of $k$ and an enhancement in the Seebeck coefficient around the Fermi energy for AlSi$_6$P-1 and AlSi$_6$P-3. This is attributed to a larger band gap and a symmetry breaking (asymmetry property) appearing in the PDOS of these two structures, which lead to a high figure of merit compared to 
pure silicene and AlSi$_6$P-1.

\subsection{Optical properties}

A knowledge of the optical properties of silicene is
very important for both spectroscopic studies of the 2D materials and optoelectronic
applications. The frequency-dependent optical characteristics of the 2D silicene are derived from the
dielectric tensor using independent-particle approximation (IPA) implemented in the QE code. 
The two independent in-plane (E$_{\parallel}$) and out-of-plane (E$_{\bot}$) components of the dielectric tensor can be calculated in the QE.
\begin{figure}[htb]
	\centering
	\includegraphics[width=0.5\textwidth]{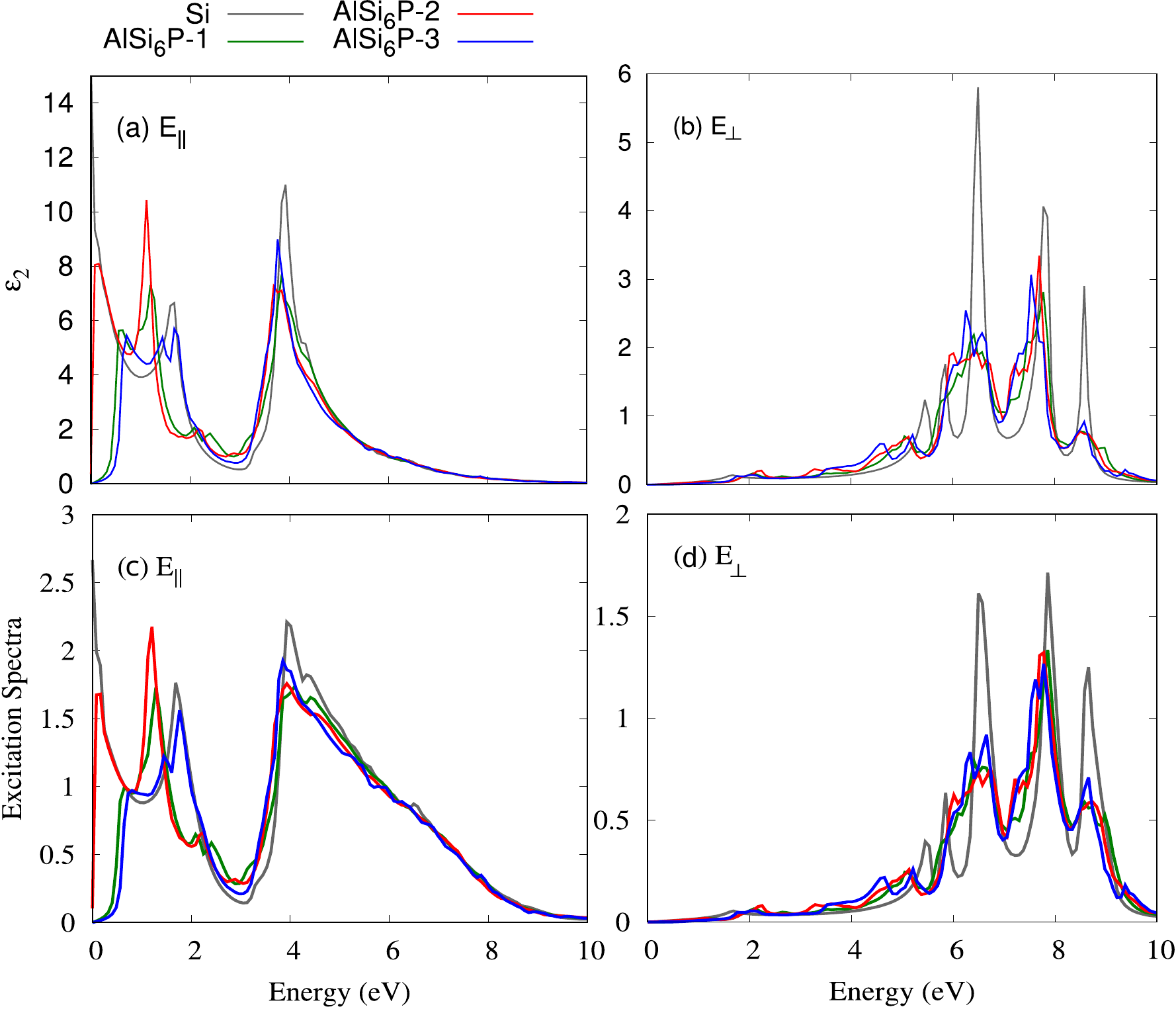}
	\caption{Imaginary part of dielectric function, $\varepsilon_2$, and excitation spectra, $k$, in the case of electric field that is parallel (E$_{\parallel}$) and perpendicular (E$_{\bot}$) to the 2D silicene and AlSi$_6$P structures.}
	\label{fig06}
\end{figure}
To calculate the dielectric tensor, the real, $\varepsilon_1$, and the imaginary, $\varepsilon_2$, parts of the dielectric functions, a sufficient number of unoccupied states above the Fermi level have to be used. In the the E$_{\parallel}$ calculations, the direction of the electric field is chosen to be parallel to the plane of the silicene nanosheet, whereas for the E$_{\bot}$ the direction of the electric field is chosen to be incident perpendicular to the plane of the silicene nanosheet.

First, we study the imaginary part of dielectric function and the excitation spectra, which are shown in \fig{fig06} for both directions of polarization of the electric fields. The peak structures in $\varepsilon_2$, and $k$ are related to inter-band transitions.
In the case of pure silicene, two main peaks are found in $\varepsilon_2$, and $k$ for E$_{\parallel}$. The two peaks are connected to $\pi\text{-}\pi^{*}$ transitions at energy $1.68$~eV along the M-K, and the $\sigma\text{-}\sigma^{*}$ transition at $3.92$~eV along the M-K and $\Gamma$-K directions. In the case of E$_{\bot}$, similar
to graphene, inter-band transitions are observed for pristine silicene except the
transitions here occure below $10$~eV. The results for peak position and intensity of pure silicene very well agree with results in the literature \cite{JOHN2017307, ABDULLAH2021114644}.  

The optical spectra are modified with the Al and P dopant atoms.  
It is interesting to see that the $\varepsilon_2$, and $k$ for AlSi$_6$P have structures that are strong at low energy for both the E$_{\parallel}$ and E$_{\bot}$. The low energy range is located at the visible and the UV regions. 
The strong peak at low energy can be related to the band structure of AlSi$_6$P. First, the opening of a band gap at the K point. Second, the gap at the $\Gamma$ point in the band structure for all AlSi$_6$P decreases due to the presence of attractive interaction between Al and P atoms.

\begin{figure}[htb]
	\centering
	\includegraphics[width=0.5\textwidth]{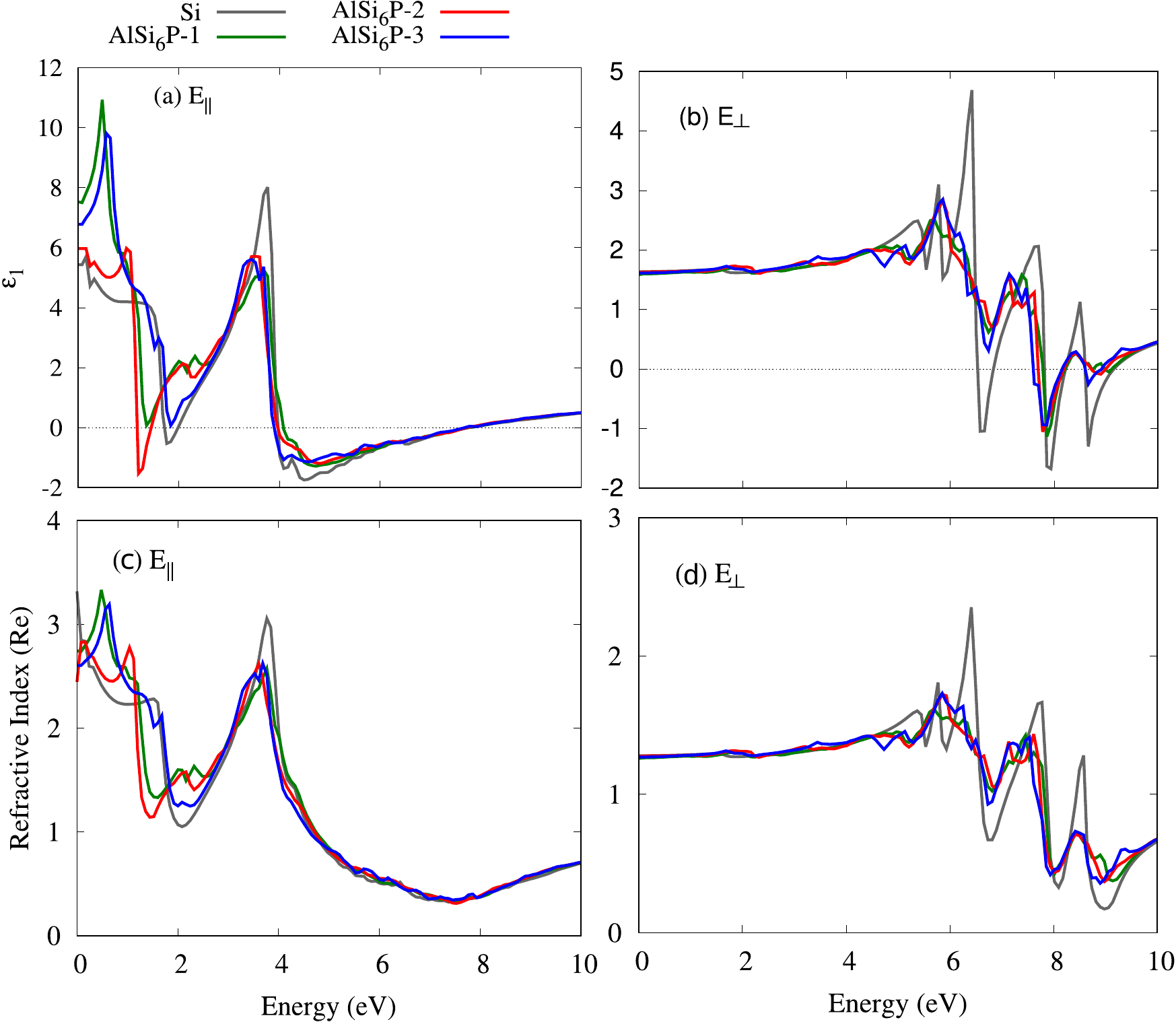}
	\caption{Real part of dielectric function, $\varepsilon_1$, and Refractive index, $n$, in the case of electric field that is parallel (E$_{\parallel}$) and perpendicular (E$_{\bot}$) to the 2D silicene and AlSi$_6$P structures.}
	\label{fig07}
\end{figure}

Similarly, the real part of the dielectric function, $\varepsilon_1$, and the refractive index, $n$, for pure silicene and AlSi$_6$P structures have almost the same qualitative behavior for both polarizations of the electric field as is presented in \fig{fig07}.

The static values of the real part the dielectric function, $\varepsilon_1(0)$, of the pure silicene are $5.43$ and $1.63$ for E$_{\parallel}$, and  E$_{\bot}$ respectively, which are in agreement
with the observation \cite{C4RA07976K}. The static value of the refractive index of pure silicene, $n(0)$, is $3.31$ and $1.27$ for E$_{\parallel}$, and  E$_{\bot}$, respectively, agreeing well with \cite{doi:10.1063/1.5062764}. The $\varepsilon_1$ is an indication of the degree of polarization. The greater the degree of polarization, the greater the value of $\varepsilon_1$ to be seen. It is interesting to see that all three AlSi$_6$P structures have higher value of $\varepsilon_1(0)$ indicating the higher degree of polarization of the atomic configurations. The higher value of $\varepsilon_1(0)$ refers to the attractive interaction between the Al and P dopant atoms. 
On the other hand, the value of $n(0)$ of AlSi$_6$P is decreased in the presence of the Al and P atoms
corresponding to the increase of the speed of light in these structures. The $n$ spectra is quite in agreement with the $\varepsilon_1$ because the increase in the degree of polarization of an atomic configuration should lead to an increase of the speed of light in the structures. This is what we see from 
the $\varepsilon_1$ and the $n$ spectra.

\begin{figure}[htb]
	\centering
	\includegraphics[width=0.5\textwidth]{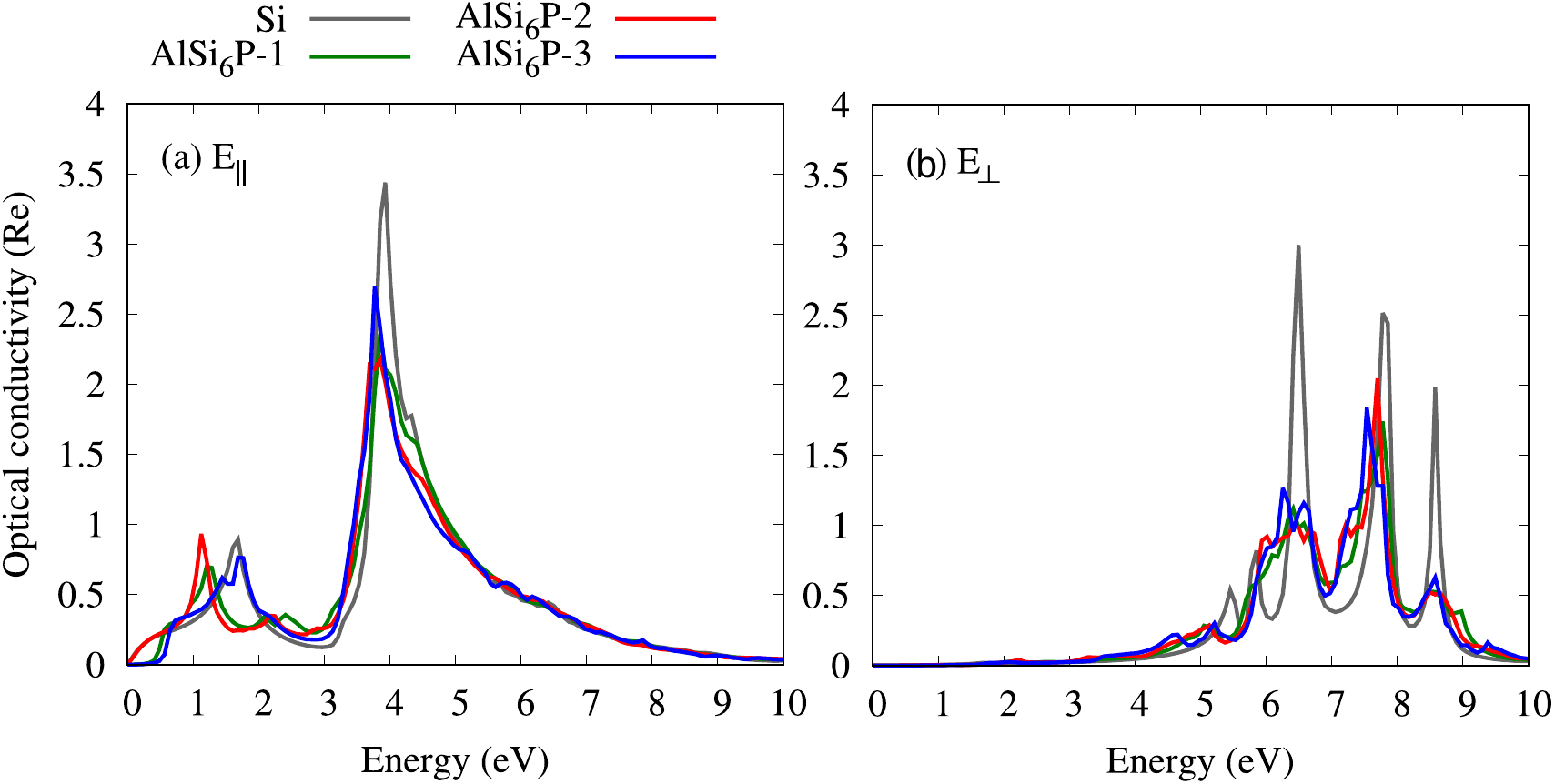}
	\caption{Real part of optical conductivity in the case of electric field that is parallel (E$_{\parallel}$) and perpendicular (E$_{\bot}$) to the 2D silicene and AlSi$_6$P structures.}
	\label{fig08}
\end{figure}

We next study the optical conductivity of pure silicene and the three configurations of AlSi$_6$P structure presented in in \fig{fig08}, where the real part of the optical conductivity is shown for 
both direction of electric field polarization. The optical conductivity is a measure of electrical conductivity for an alternating field. One can easily recognize the optical gap from optical conductivity spectra of AlSi$_6$P-1 and AlSi$_6$P-3 at low energy in the case of E$_{\parallel}$. In addition, a red shift in leftmost peak of the optical conductivity is found for the AlSi$_6$P-1 and AlSi$_6$P-3 in the case of E$_{\parallel}$. The red shift of the peaks is caused by the decreased energy spacing between the $\pi$ and the $\pi^*$ states, and the $\sigma$ and the $\sigma^*$ states along the $\Gamma$-M and the M-K directions 

\section{Concluding remarks}\label{Sec:Conclusion}

We have performed DFT calculations to investigate structural, electronic, mechanical, thermal and optical characterstics of pure buckled silicene and Al-P codoped silicene structures by taking into account the interaction effect between dopant atoms.
It was found that the substitutions of Al and P atoms do not preserve the linear energy dispersion due to breaking of the inversion symmetry, besides the position and the Fermi level depends on the attractive interaction between the substitution atoms. 
In addition, a high Seebeck coefficient and electronic thermal conductivity were found in the presence of the attractive interaction. It should be mentioned that the Al-P codoped silicene has a weaker mechanical reponse compared to the pure silicene structure.
With the presence of the attractive interaction between the dopant atom, the peaks in the dielectric function and the optical conductivity spectra are red-shifted irrespective of the nature of polarization of the incident field.

\section{Acknowledgment}
This work was financially supported by the University of Sulaimani and 
the Research center of Komar University of Science and Technology. 
The computations were performed on resources provided by the Division of Computational 
Nanoscience at the University of Sulaimani.  
 


\end{document}